\documentclass{IEEEtran4PSCC}

\ifCLASSINFOpdf
 \usepackage[pdftex]{graphicx}
\else
\usepackage[dvips]{graphicx}
\fi
\usepackage[cmex10]{amsmath}
\newcommand*{\figuretitle}[1]{%
 {\centering% <-------- will only affect the title because of the grouping (by the
 \textbf{#1}%    braces before \centering and behind \medskip). If you remove
 \par\medskip}%   these braces the whole body of a {figure} env will be centered.
}

\usepackage{xcolor} %% for textcolor
\usepackage[caption=false]{subfig}

\usepackage[ruled,vlined,linesnumbered]{algorithm2e}

\makeatletter
\let\old@ps@headings\ps@headings
\let\old@ps@IEEEtitlepagestyle\ps@IEEEtitlepagestyle
\def\psccfooter#1{%
  \def\ps@headings{%
    \old@ps@headings%
    \def\@oddfoot{\strut\hfill#1\hfill\strut}%
    \def\@evenfoot{\strut\hfill#1\hfill\strut}%
  }%
  \def\ps@IEEEtitlepagestyle{%
    \old@ps@IEEEtitlepagestyle%
    \def\@oddfoot{\strut\hfill#1\hfill\strut}%
    \def\@evenfoot{\strut\hfill#1\hfill\strut}%
  }%
  \ps@headings%
}
\makeatother

\begin{document}
\title{Real-time Anomaly Detection and Classification\\ in Streaming PMU Data} 

% \author{
% \IEEEauthorblockN{Christopher Hannon, Dong Jin %[Student(s) to be added at a later stage]
% }
% \IEEEauthorblockA{%Computer Science Department\\ % save space
% Illinois Institute of Technology\\ Chicago, IL 60616 United States\\
% %\{lokhov, vuffray, deepjyoti, chertkov\}@lanl.gov}
% }
% \and
% \IEEEauthorblockN{Deepjyoti Deka, Andrey Y. Lokhov%, Marc Vuffray %[Student(s) to be added at a later stage]
% }
% \IEEEauthorblockA{%Theoretical Division\\ % save space
% Los Alamos National Laboratory\\ Los Alamos, NM 87544 United States\\
% %\{lokhov, vuffray, deepjyoti, chertkov\}@lanl.gov}
% }
% }

\author{\IEEEauthorblockN{Christopher Hannon\IEEEauthorrefmark{1}\IEEEauthorrefmark{2}\IEEEauthorrefmark{3},
Deepjyoti Deka\IEEEauthorrefmark{1},
Dong Jin\IEEEauthorrefmark{3}, 
Marc Vuffray\IEEEauthorrefmark{1} and
Andrey Y. Lokhov\IEEEauthorrefmark{1}}
% Template-style formatting
% \IEEEauthorblockA{\IEEEauthorrefmark{1} Theoretical Division\\
% Los Alamos National Laboratory,
% Los Alamos, NM 87544\\ \{lokhov,vuffray,deepjyoti,chertkov\}@lanl.gov}
\IEEEauthorblockA{\IEEEauthorrefmark{1} Theoretical Division,
Los Alamos National Laboratory,
Los Alamos, NM 87544}
%\{lokhov,vuffray,deepjyoti,chertkov\}@lanl.gov
\IEEEauthorblockA{\IEEEauthorrefmark{2} Center for Nonlinear Studies,
Los Alamos National Laboratory,
Los Alamos, NM 87544}
\IEEEauthorblockA{\IEEEauthorrefmark{3} Computer Science Department, 
Illinois Institute of Technology, Chicago, IL 60616}
}

\maketitle

\begin{abstract}
Ensuring secure and reliable operations of the power grid is a primary concern of system operators. Phasor measurement units (PMUs) are rapidly being deployed in the grid to provide fast-sampled operational data that should enable quicker decision-making. This work presents a general interpretable framework for analyzing real-time PMU data, and thus enabling grid operators to understand the current state and to identify anomalies on the fly. Applying statistical learning tools on the streaming data, we first learn an effective dynamical model to describe the current behavior of the system. Next, we use the probabilistic predictions of our learned model to define in a principled way an efficient anomaly detection tool.
% design an efficient anomaly detection tool that identifies events by measuring deviations from the probabilistic predictions of our learned model.
Finally, the last module of our framework produces on-the-fly classification of the detected anomalies into common occurrence classes using features that grid operators are familiar with. We demonstrate the efficacy of our interpretable approach through extensive numerical experiments on real PMU data collected from a transmission operator in the USA. % and release the associated code-base that can be directly tested on other systems due to the intrinsic transferability of our framework.
\end{abstract}
% Use this to place sponsorships
\thanksto{The work was supported by funding from the U.S. DOE as part of the DOE Grid Modernization Initiative, and by the Laboratory Directed Research and Development program of Los Alamos National Laboratory under project number 20190351ER.}

%%%%%%%%%%%%%%%%%%%%%%%%%%%%%%%%%%%%%%%%%%
           %
\section{Introduction}
\begin{figure}%[b]
\centering
\figuretitle{Example Anomalies}
\includegraphics[scale=0.3]{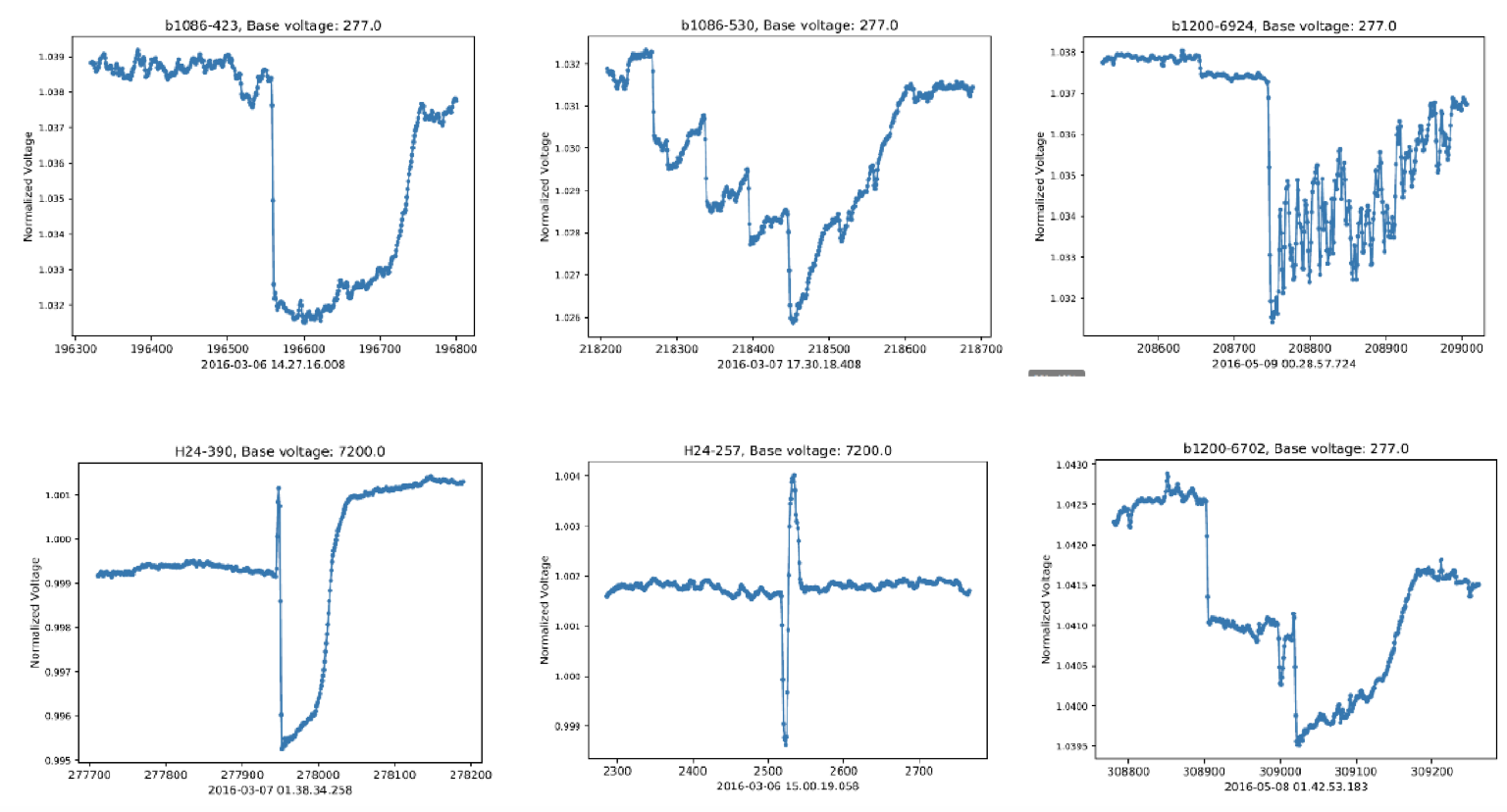}
\caption{Example of anomalies detected in voltage signal of a US Independent System Operator.}
\label{fig_anom_ex}
\end{figure}

Traditional supervisory control and data acquisition (SCADA) systems provide information regarding the system state at the order of seconds to the operator. However, such fidelity, considered appropriate in prior decades, is not sufficient to observe or predict disturbances at faster time-scales that are increasingly being observed in today's stochastic grid \cite{hossain2011investigation}. %maintenance is a complex task that has been approached in various ways for decades, with no approaches having the level of sophistication and access to technology that we are currently presented with. 
To provide more rapid measurement data, phasor measurement units (PMUs) have gained widespread deployment.
PMUs \cite{pmu} are time-synchronized by GPS timestamps and collect measurements of system states (Eg. voltage, frequency, currents) multiple times per second at rates between 30 and 240 Hz. Thereby, if efficient methods for PMU analysis are developed, they can help in obtaining a more holistic understanding of the state of the power grid. Real-time detection of changes in the power system can have multiple benefits to grid security. Among others, it can enable corrective control actions by operators that can prevent grid failures from devolving into cascading blackouts \cite{ghanavati2015identifying}. Together with event-detection, the problem of event identification/classification is equally important as it helps determine probable causes and follow-up actions if necessary. Figure~\ref{fig_anom_ex} shows several voltage anomalies measured by a PMU system in a US transmission grid. While PMUs have gained increasing deployment in the modern power system, the use of their data reporting in real-time operations has not yet realized full adoption. The primary use of PMU data so far has been for offline event discovery and post-event analysis \cite{PMU_loc}.

The overarching goal of this research is to create a \emph{real-time} approach for facilitating a remote analysis of individual components in the grid for monitoring, detection, and classification of anomalies. The implementation of our framework is fully automated, user-friendly and crucially \emph{interpretable}. Any time the warning flag about the system state is raised, our methods ensure that the operator is able to see quantitative details of the prediction and classification sub-tasks in real-time for validation. %This provides grid operators insight into the decision-making and can help them identify problematic future occurrences in the system. 

A principled way to achieve the interpretability of predictions consists of connecting data with the physical properties of the system. This physics-informed approach has been proven successful when a reasonable system model under given conditions is known, such as the recovery of linearized swing equations in transmission-level power grids from PMU observation \cite{lokhov2018online}. However, often even the functional form of the model is not known, or the conditions under which it holds can not be verified \emph{a priori}; hence, it becomes desirable to use a more flexible data-driven approach that is based on, but also extends the physics-based models commonly used in modeling the dynamics of power grids.   

The data-driven framework proposed in this work is principally based on the theory of robust statistical machine learning \cite{wainwrightjordan}. In particular, we learn the parameters of a general stochastic vector auto-regressive (VAR) process to model the effective relationship between measurements collected. This multivariate approach, updated at regular intervals, makes near-term predictions of future trajectories for the data. The anomalies are then identified based on statistically significant deviations in the incoming data from the ensuring predictions. Finally, we design a decision-tree based classification tool into our framework to distinguish the identified anomalies into different interpretable categories. In what follows, we discuss the application of our learning framework on real PMU data-feed collected from a US transmission system operator (ISO).

Compared to uni-variate data analysis and other general-purpose machine learning techniques such as neural networks, our multivariate approach coupled with decision tree-based classification shows superior performance for both event detection and classification in the available data-set. Even more importantly, our approach does not involve tuning or fitting of thresholds and hyper-parameters specific to a particular system, and hence is intrinsically transferable from one real system to another. 
%In the accompanying online repository \cite{our_code}, we provide the code associated with our framework for testing and validation by interested users. 

The rest of the paper is organized as follows: Section~\ref{sec-related} presents related work, %Section~\ref{sec-framework} outlines the statistical learning-based framework for PMU data stream modeling and applications. 
Section~\ref{sec-approach} presents the implementation of our machine learning framework and the procedure for setting proper parameter values in the statistical model. 
In Sections~\ref{sec-anom} and ~\ref{sec-classification}, the anomaly detection and classification application using the implementation of the framework is presented and evaluated. 
Finally, the implications of the framework with proposed future work are discussed in Section~\ref{sec-discuss}. 

\section{Related Work}
\label{sec-related}
Real-time use of PMUs is a growing field of research, though its practical implementation has been sporadic. The approach presented in this paper provides a statistical learning framework for modeling the multivariate stream of raw data from a single PMU that measures frequency, and complex-valued voltage, current, and power measurements in real-time. This step is motivated by non-trivial dependence between single-variable measurements at the same location, as observed in later sections. Conversely, in prior work \cite{pmu_small, pmuOther}, statistical tests with sliding windows are designed to detect anomalies in single variables. In \cite{PMU_loc}, the authors use a rule-based approach for detecting changes in each data variable independently. The approach utilizes off-line analysis of data rather than the real-time approach presented in this work.
In \cite{smartMeter}, the authors focus on detecting and classifying smart meter anomalies using unsupervised machine learning techniques, including neural network, and projection-based methods.
The low data-rate of smart meters makes it non-trivial to extend the techniques to real-time PMU data.
In \cite{PMU_ensemble}, the authors develop an ensemble-based method that combines multiple detection algorithms to create a more robust detection mechanism. In \cite{PMU_decision}, the authors use a decision tree to classify line events from PMU data. While our work utilizes decision trees as well, due to their interpretability, the detection methods used makes it different from the existing work.% due to the detection methods. 
A guide for signal processing based techniques is developed in \cite{nrel_long}. The authors use the voltage phase angle difference measured by PMUs with a reference bus to determine power system events. 
In contrast, we model the PMU data stream individually within its local context for use in real-time detection. In \cite{PMU_comms}, the authors use K-nearest-neighbor (KNN) and Singular Spectrum Analysis (SSA) combined with fog computing which moves processing to edge devices. In future work, we intend to evaluate our modeling methods on multiple PMU data streams and incorporate edge computing paradigms.

Overall, the approach presented in this paper attempts to generalize the problem of anomaly detection: detection and scoring of anomalies in a principled manner based on an effective model that is learned on the fly; well-defined data-driven approach to selecting hyper-parameters; capturing the interaction of variables due to underlying system properties; and working in the invariant spaces of probabilistic scores which ensures transferrability of methods. 
% Additionally, our modeling approach is resolution independent.
% per variable, 
%\textcolor{blue}{\bf A: good review of existing work. However, can we say more? Like, their methods are not transferable, depend on the choice of thresholds, maybe crucially depend on some system properties... Because so far the only difference highlighted is the single vs. multi variables analysis. But from our results we see that one variable is often enough...}

\section{Approach}
\label{sec-approach}
To analyze the PMU measurements collected from the grid during normal ambient operation, we model the observed values using a dynamical model. Once a suitable effective descriptive model is obtained, we can gain insight into the state of the system and changes made in it. We start by describing the measurements available from a US ISO for analysis.

\subsection{Data Description and Preliminary Analysis}
%\textcolor{blue}{\bf A: Deep, what is the final status of this data? Just to be sure: it is OK to publish results on it as long as the name is not revealed? Normally it was the case, but the NDA made by Scott might have expired, at least Dan Bienstock was renewing their NDA. On the other hand, he was OK to share data with LANL, and Misha is on that paper. So now I'm a bit confused about the status. I think it's OK, but let's make sure not to get into trouble.}

\begin{figure}%[b]
\centering
\figuretitle{Correlation of PMU Data Stream Variables}
\includegraphics[scale=0.56]{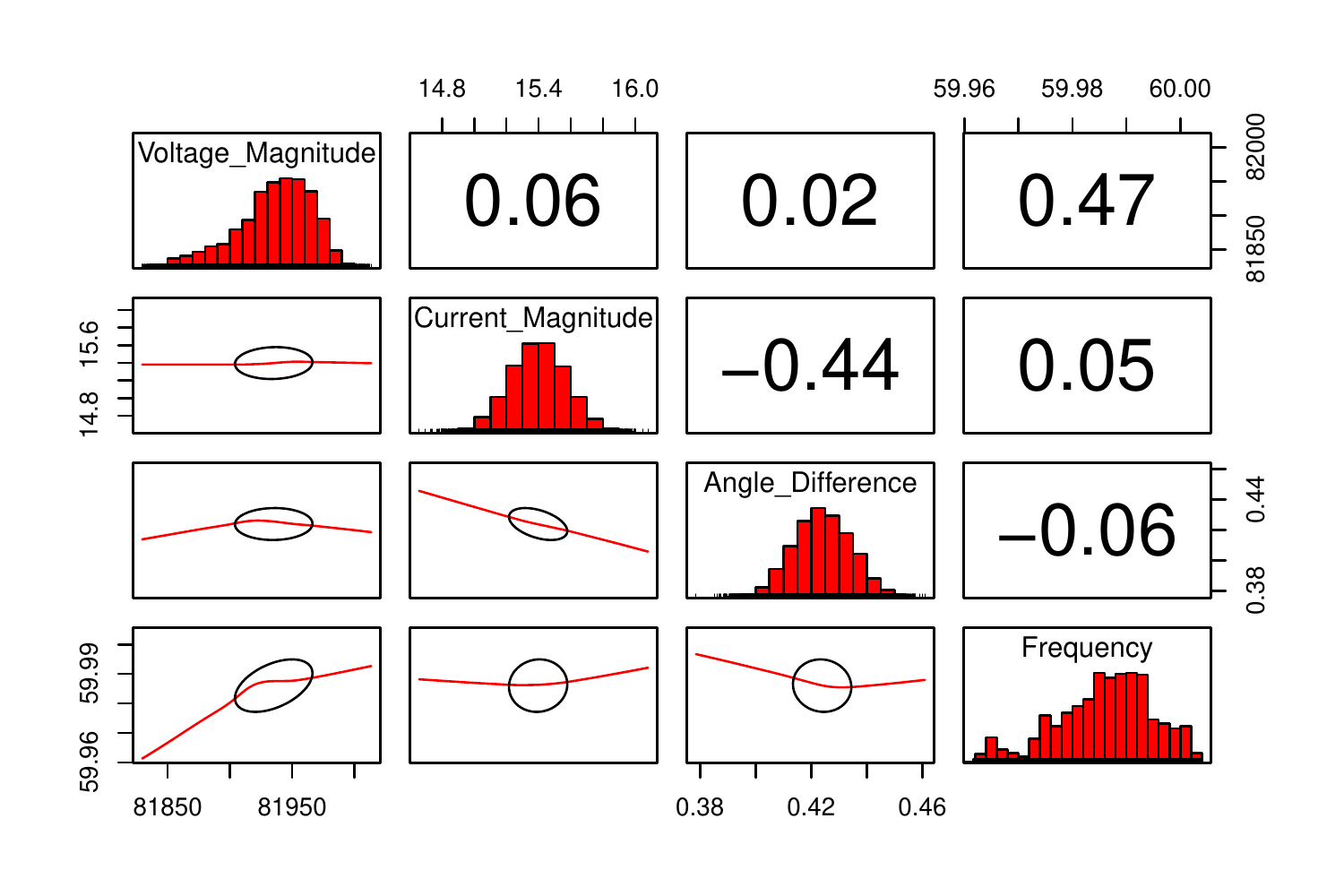}
\caption{Correlation of variables over 90 minutes of stable operation. Bivariate scatter plots are shown below the diagonal as LOESS lines, histograms of the distribution of values on the diagonal, and the Pearson correlations above.}% the diagonal. }
\label{fig_corr_some}
\end{figure}

%The electric power grid is organized into a network of nodes called buses (which include connected loads and generators) and edges which are lines that connect the buses. Measurements are collected at the components to monitor the state of the system as well as the buses where energy is transferred. These include measurements collected by phasor measurement units (PMUs) that are time-synchronized by GPS. 
To provide a foundation for our statistical model, we use time-stamped vector-valued data collected from a single PMU on the transmission system of a US ISO. In this study, we use $87$ days of data, collected at 30Hz. Due to the underlying physics of the grid, there is an inherent relationship between the power variables useful to power operators in the data-stream. %, listed in Table \ref{tab:my_label}.
\if 0
\begin{table}[ht]
 \centering
 \caption{Data stream variables inferred from the PMU measurements.}
 \begin{tabular}{|l|l|}
 \hline
  Voltage Magnitude & Measured in volts \\ \hline
  Voltage Angle & Measured in degrees \\ \hline
  Current Magnitude & Measured in Amperes \\ \hline
  Current Angle & Measured in degrees \\ \hline
  P & Real or active power in Megawatts (MW) \\ \hline
  Q & Reactive power in megavolt-amperes (MVAR)\\ \hline
  S & Complex Power in megavolt-amperes (MVA), \\ 
  & Apparent Power is magnitude of (S) \\ \hline
  PF & Power factor is the ratio of real power to \\ 
  & apparent power\\ \hline
  % DF/DT & Rate of change of frequency \\ \hline
  Frequency & Measured in Hertz\\ \hline
 \end{tabular}
  \label{tab:my_label}
\end{table}
\fi
The power triangle relates active power (P), reactive power (Q), Complex Power (S), and phase angle difference $\phi$.  
Coupling with voltage, current and frequency, there are many potential variables at hand to model the system.
%Hence, some of the data streams are redundant. 
Using all the information available is not optimal from the information storage and real-time data processing points of view; additionally, exact strong correlation relations between some of the variables may artificially complicate the learning of an effective model. Therefore, we limit the data variables used in our model %that we wish to model 
to voltage, current, the sine of the angle difference of voltage and current and frequency. All other quantities can be straightforwardly estimated from these four if required.
Figure~\ref{fig_corr_some} shows the empirical correlation between the variables used in our model. On the diagonal of the figure, the histograms of the data for each plot are shown. The plots below the diagonal contain the bivariate distribution ellipses to 95\% confidence of the values along with their locally estimated scatter plot smoothing (LOESS) lines. Above the diagonal, the Pearson correlation coefficients are listed. 
%The Pearson correlation coefficient, $\rho$ is calculated for a pair of random variables $(X,Y)$, covariance function \texttt{cov}, and standard deviations $\sigma_X$ and $\sigma_Y$ of $X$ and $Y$ respectively by:

\if 0
\begin{equation}
 \rho_{X,Y} = \frac{\texttt{cov}(X,Y)}{\sigma_X\sigma_Y } 
 \label{fomula_corr}
\end{equation}
\fi
%\textcolor{blue}{\bf A: People can look up the definition of Pearson correlation in Widepedia, so we can remove the last sentense above.}

A simple exploratory analysis of the Figure~\ref{fig_corr_some} indicates that the sine of the angle difference is correlated to the magnitude of current and that voltage magnitude is correlated to the frequency of the system. This observation motivates the need to take the multivariate nature of the signal into account when constructing the effective dynamical model of the system, which is one of the main features of our method. 

\subsection{Data Modeling}
To model the PMU data stream, we design a multivariate approach for capturing the interaction between variables. The interaction is expected due to the underlying physics of the electric power grid. In general, the dynamics may have an arbitrary non-linear functional complexity; however, a common approach consists of considering a liner effective model linearized around a current operating point, which is typically valid over short time scales. Hence, in what follows we use the four identified variables in a vector autoregressive model (VAR) which relates the system's current state to time-lagged values of previous states via dynamical state matrices and an error term.
% The vectorized form enables the current state of each variable to depend on the previous states of all variables. %%% A: This statement is a bit confusing and unnecessary

A VAR(p)-process is defined consisting of $K$ endogenous variables $y_t = (y_{1t},...,y_{kt},...,y_{Kt})$ for $k = 1,...,K$ as: 
\begin{equation}
 y_t = c + A_1y_{t-1} + ... + A_py_{t-p} + u_t \ ,
 \label{formula_var}
\end{equation}
with $A_i$ as $K \times K$ coefficient matrices for $i = 1,...,p$ and $u_t$ is a $K$-dimensional random noise process with expected mean, E$(u_t) = 0$ and time invariant white noise positive definite covariance matrix $E(u_tu_t^\top) = \Sigma_u$ \cite{vars2}. In the case of an independent Gaussian noise, the coefficients of a VAR model can be optimally estimated using least-squares applied to the rewritten equations for each $k \in K$ variable separately \cite{lokhov2018online,lokhov2018cdc}.

\begin{figure}%[]
\centering
\figuretitle{PMU Data Stream and Standardization}
\includegraphics[scale=0.45]{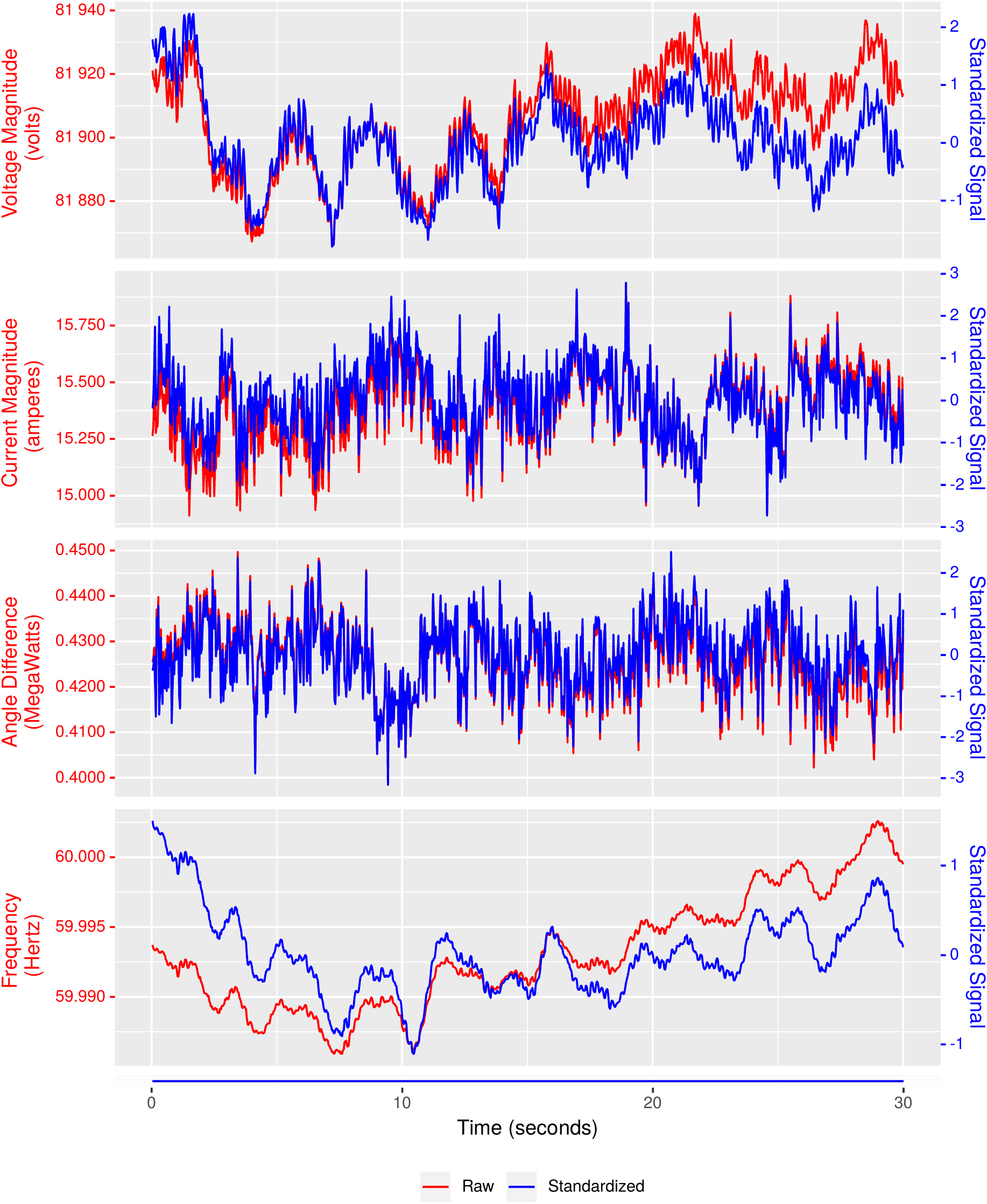}
\caption{The PMU data stream is shown in red for each variable over 30 seconds, with scale on left y axis. The standardized signals are shown in blue using the right side y axis. The standardization is achieved by linear de-trending and dividing by the standard deviation of the population over a chosen period of time, the training data size for the VAR model.}% (\textcolor{blue}{\bf XXX minutes in the tests presented below}).}
\label{fig_raw_30sec}
\end{figure}

Since different variables have different scales, it is convenient to first pre-process and adequately format the data for the statistical learning model. First, each signal is individually de-trended over the period the model learns on. To normalize, we then divide the data interval by the standard deviation, $\sigma_k$ for each variable $k \in K$. Thus our transformed data is measured in standard deviations from the linear trend-line and original units such as volts, amperes are removed. Figure~\ref{fig_raw_30sec} shows an example of regular raw data signals next to the standardized signals. A VAR process is then fitted to the pre-processed data. The R language VAR implementation is used \cite{vars1} for analysis. Notice that it is immediate to transform the standardized data back to physical scale and dimension using the saved values of the linear trend and value of the standard deviation. The standardization parameters are constantly adjusted to incorporate new incoming data in real-time.

\subsection{Hyper-parameter selection in VAR}
The VAR estimation process includes the following input hyper-parameters: (a) the time resolution we want to model; (b) $\tau$, the
% number of samples %%%A: should be called time duration from what follows
length of the time period over which the VAR coefficients are trained; and (c) $p$, the number of lag terms included in the model. Below, we describe a principled empirical procedure for selecting the right values of each of the hyper-parameters for a given data set.

\paragraph{Model resolution} The time resolution hyper-parameter is not data-related, but instead is intrinsically tied to the time scale of the application of interest. For example, detection of forced oscillations occurring at the scale of seconds would require much high data resolution compared to slow-moving changes such as transformer aging \cite{transformers} %\textcolor{red}{\cite{mani}} 
which is the process that may take weeks or even months. In this study, we are interested in real-time detection of anomalies on the maximum scale of seconds; we use a 0.5-second prediction interval that enables detection of real-time changes in the system. To fit the model using this resolution, the data should be appropriately coarse-grained to scale. In the case of a given data set, the data is processed by taking the rate of data transmission, i.e., 30 Hertz and dividing by the granularity (0.5) seconds and taking the mean of those points.

\paragraph{Training data size} Next, we determine the selection of training data length, $\tau$. For a stable model, more data (larger $\tau$) leads to a better model fit. However, the physical parameters in an electric power grid tend to change throughout the day due to changing weather, control settings, or consumption and generation. Hence, we do not expect our model to be stable for very long periods of time; the effective description should drift with time as well. Moreover, our requirement of real-time data processing means that we would like to minimize the number of data points to construct a model on the fly. Hence, an ideal amount of training data $\tau$ will balance the training error caused by a finite number of learning samples and the model misspecification error due to the naturally changing effective model of the system, at the same time keeping the computational complexity of the learning procedure low. Below, we describe a self-consistent data-driven approach to set the value of $\tau$ that would satisfy these requirements. 

\begin{figure}%[b] % figure margins 6x8 in
\centering
\figuretitle{Accuracy of Fitting VAR(1) Model}
\includegraphics[scale=0.77625]{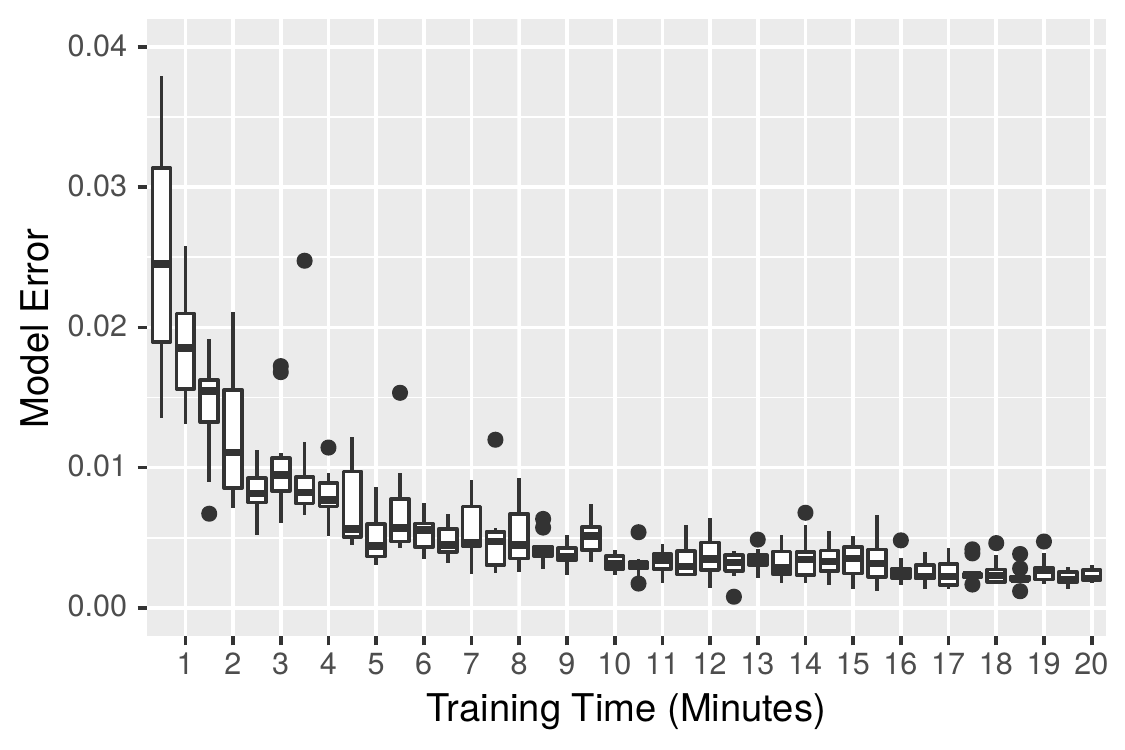}
\caption{Estimation of the training error. Data is synthesized from VAR(1) models trained on the available PMU data. VAR(1) models are then fit on the same amount of synthesized data. Plotted is the average difference between the trained $A$ state matrix and the re-trained $\hat A$ coefficient matrix. Their distributions are plotted for training times varying from 0.5 minutes to 20 minutes over 50 independent data intervals. 
}
\label{fig_retrain}
\end{figure}

How can one estimate the training error given that the ground truth model is unknown? To tackle this challenge, we propose to learn a trial VAR(1) model from data and then use it to synthesize data; this model will then serve as ground truth in this synthetic experiment. From the synthesized data, we then retrain a VAR(1) model and calculate the difference between the original and the inferred models. A larger difference would show that there is an insufficient amount of data to accurately reconfigure the VAR model. In Figure~\ref{fig_retrain}, data is synthesized from VAR(1) models trained on PMU data from $\tau = [0.5, 20]$ minutes at half-minute intervals. VAR(1) models are then retrained on the synthesized data. The error is computed using the following relation defined by the average absolute per-element difference $D_1$: 
\begin{equation}
 D_1 = \frac{\sum_{ij} |\hat A_{ij} - A_{ij}|} {K^2}
 \label{formula_l1}
\end{equation}
Here, $A$ is a reference VAR(1) coefficient matrix and $\hat A$ is the VAR(1) coefficient matrix fitted on data simulated using $A$. Figure~\ref{fig_retrain} shows that the error rapidly decreases to an average of about 0.005 at about 8 minutes, after which the decay becomes gradual. This illustrates that the typical dynamic state matrix for the VAR(1) model is reconstructed to an acceptable accuracy using the data equivalent of 8-12 minutes. 

In order to evaluate the mis-specification error arising from the natural change of the effective model, we compare the coefficient matrices estimated using two consecutive intervals:
\begin{equation}
 D_2 = \frac{\sum_{ij} |A_{ij} - B_{ij}|} {K^2}
 \label{formula_l2}
\end{equation}
Similarly to Eq.~\ref{formula_l1}, Eq.~\ref{formula_l2} computes the average absolute per-unit difference of $K \times K$ $A$ and $B$ where $A$ is the coefficient matrix from the VAR(1) model trained from $[t,t+\tau)$ and $B$ is the coefficient matrix fitted from the VAR(1) model trained from $[t+\tau, t+2\tau)$ for $\tau$ ranging from 0.5 to 20 minutes.

\begin{figure}%[b] % figure margins 6x8 in
\centering
\figuretitle{Change in Model Over Data Stream}
\includegraphics[scale=0.775]{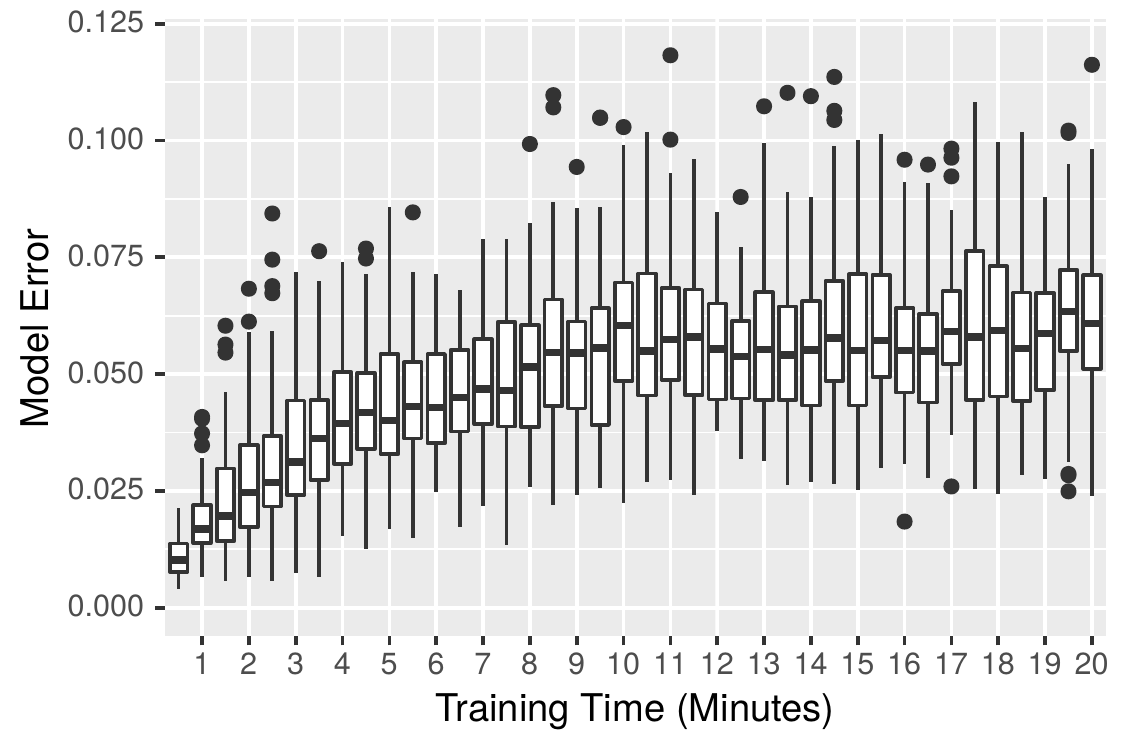}
\caption{The error is plotted over various $\tau$ values ranging from 0.5 to 25 minutes. The error is computed by Equation~\ref{formula_l2}. 
The error increases and then levels off at about 10 minutes.
}
\label{fig_mod_change}
\end{figure}

Figure~\ref{fig_mod_change} illustrates the stability of the model when $\tau$, the size of the training set, increases. As expected, over short intervals i.e., 1 minute, the model remains very stable with less than 0.025 per element change in the coefficient matrices. As the length of the model increases, the similarity of the coefficient matrices decreases. After about 10 minutes, the error plateaus around 0.05. Based on our similarity metric, it can be concluded that beyond 10 minutes, the difference in matrices encompasses the error due to the change in the effective model.

\begin{figure}%[hb!] % figure margins 
\centering
\figuretitle{Error in VAR(1) and VAR(2) Models}
\includegraphics[scale=0.65]{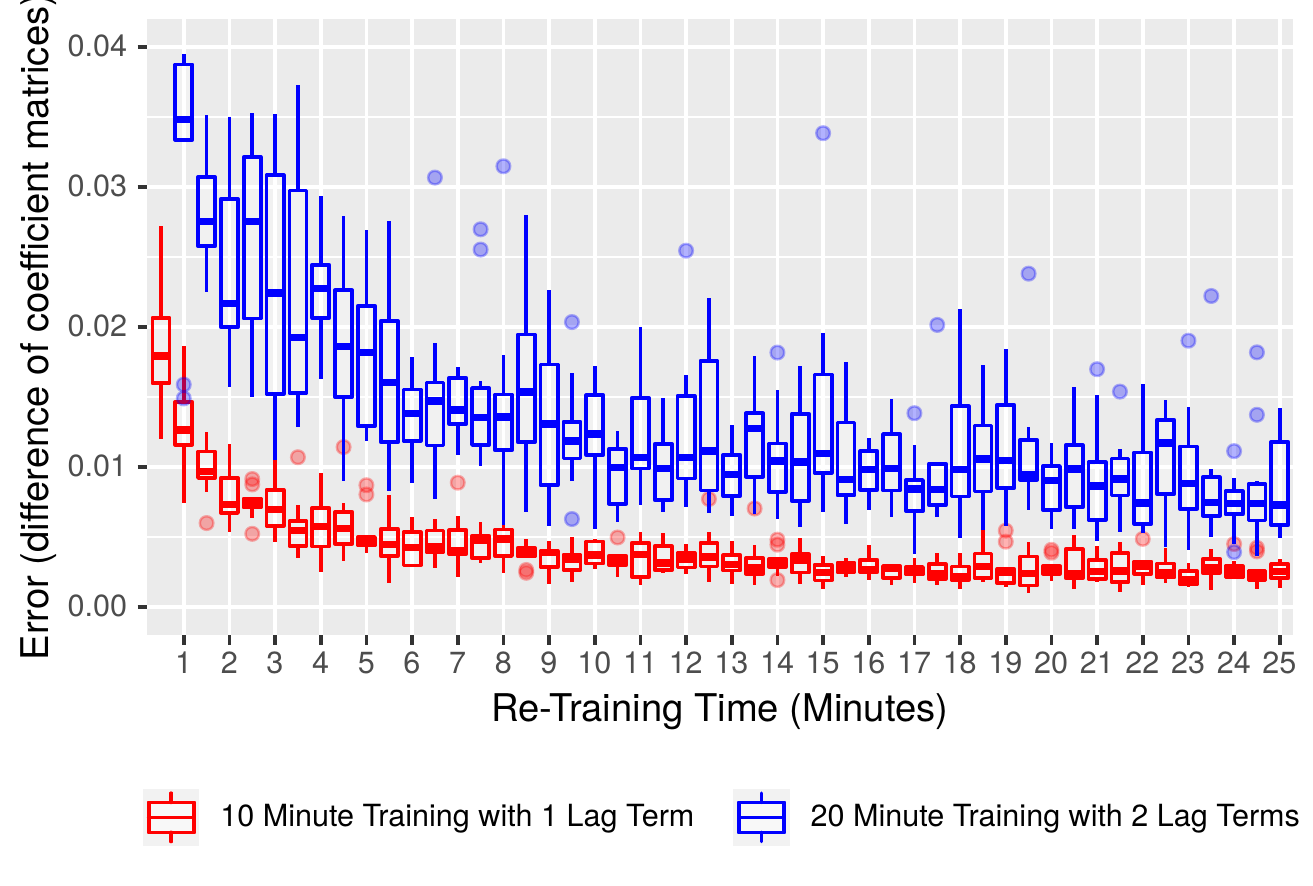}
\caption{Retraining accuracy of a VAR(1) model trained over 10 minutes of PMU data with a VAR(2) model trained over 20 minutes of PMU data. The VAR(1) model has an average per-element error of the coefficient matrix of 0.0039 when retrained on 10 minutes, while the VAR(2) model retrained on 20 minutes has an average error of 0.0087.
%\textcolor{blue}{\bf Is there a more ``vertical'' version of this figure that would fit it into a one-column format to save space?} 
}   
\label{fig_lags}
\end{figure}

The analysis of the two sources of error depicted in Figures~\ref{fig_retrain} and~\ref{fig_mod_change} shows that an increase in error due to model misspecification will take over the reduction of error due to statistical estimation. In what follows, we fix $\tau = 10$ minutes as an acceptable choice for the length of training data in the current data set.

\paragraph{Depth of the model} The final parameter of the VAR model is the number of lag terms $p$, that are used to predict the next state. The benefit of including more time lags which increases the expressivity of the model can be hampered by the decrease of the accuracy of the recovered model due to the final amount of data. In order to determine an appropriate number of lag terms to include, we analyze the impact of the number of lags on a model's retraining error using Eq.~\ref{formula_l1}. For the case of 2 lags, we compute the average per-element absolute error over both coefficient matrices $A_1$ and $A_2$. Figure~\ref{fig_lags} illustrates the error over 50 runs where $\tau$ increases from 0.5 minutes to 25 minutes. It can be seen that a VAR(2) model with $\tau = $ 20 minutes has over twice the per-element error as a VAR(1) model with $\tau = $ 10 minutes. Therefore, we can conclude that to minimize errors associated with the selected $\tau$, the inclusion of two lags is prohibitive and instead one lag is sufficient for the VAR model on the current data set.

To summarize, for the data set analyzed in this study, we utilize a VAR(1) model with a training data length of $\tau = 10$ minutes, and a 0.5-second prediction scale. The variables included in the VAR(1) model are voltage magnitude, current magnitude, sine of the phasor angle differences, and the frequency. In the next section, we use this VAR(1) model for anomaly detection and subsequent classification. For other systems and data sets, the operator may use the procedure described above to set the optimal hyperparameters of our learning framework. 

%%% A: I've moved the "coarse-graining" explanation to the resolution paragraph above

%Now that we have can recover the appropriate model fitted on the streaming data, we can score the incoming PMU data according to this probabilistic model. Power system events, referred to as anomalies in this work, can be principally detected in real-time using these scores, as explained in the following section.

%\textcolor{blue}{General comment: maybe we should say that anomaly here can refer to events as well in our data..because people also talk of event detection.}

\section{Anomaly Detection}
\label{sec-anom}
Under normal control operations as well as external perturbations, the measurements recorded in the PMU measurements change gradually. However brief or extended large changes are observed from time to time and qualified as anomalies. 
These include both operational changes such as load tap changes, as well as external perturbations such as lightning strikes. Figure~\ref{fig_anom_ex} shows a few examples of anomalies in the voltage signal of the available PMU. While large anomalies can be visually seen by the human eye in individual measurement streams, we propose a statistical approach that enables early detection of anomalies when they begin or less severe or subtle anomalies that can only be detected by considering multiple measured variables in the VAR(1) model. 

\begin{figure}%[b] % figure magrins 6x8 in
\centering
\figuretitle{Residual Distance Scores}
\includegraphics[scale=0.65]{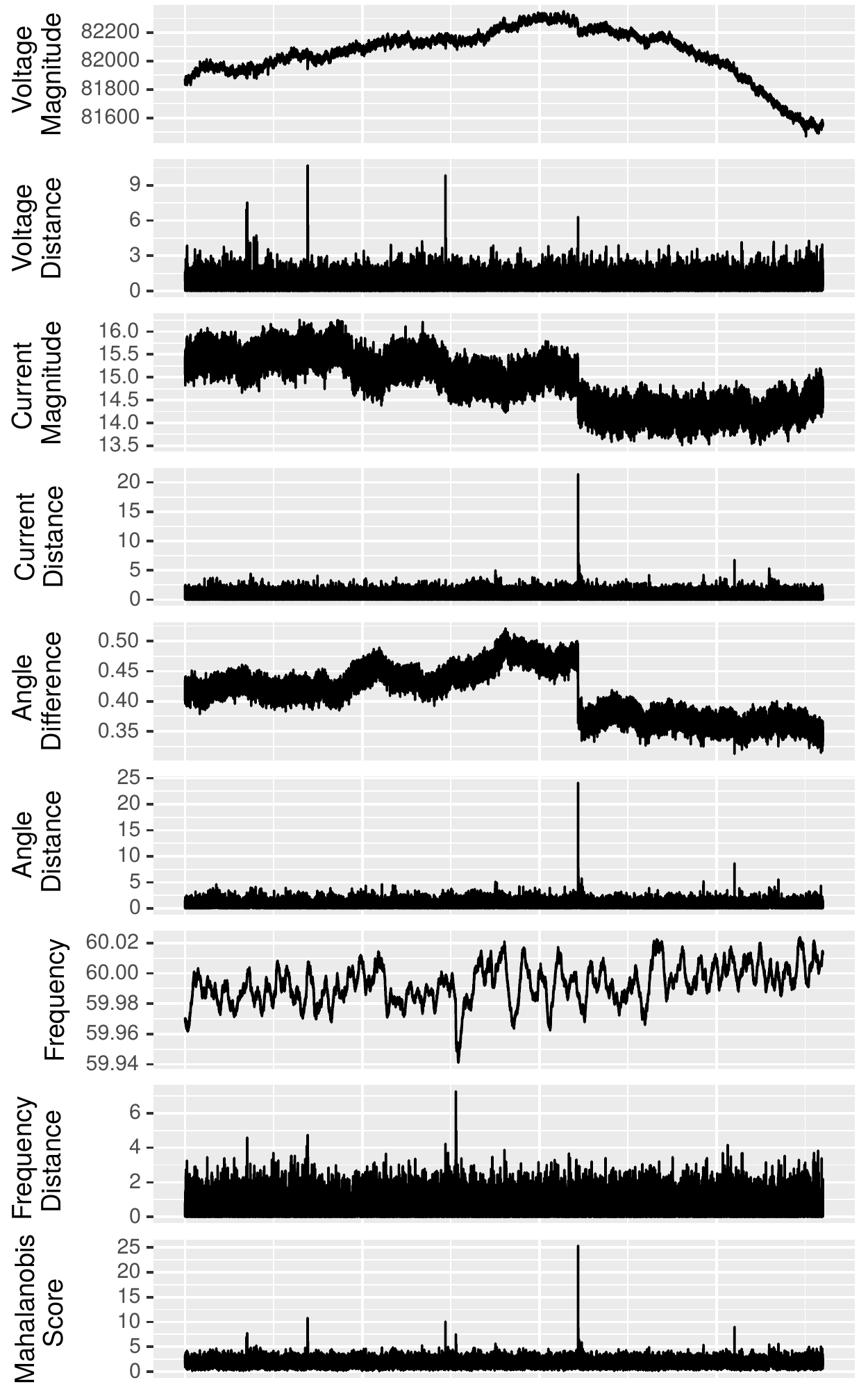}
\caption{90 Minutes of raw data and corresponding residual distance scores. The scores are measured in standard deviations.
Some anomalies are seen in the data stream by multiple variables, some by single variables, and some anomalies the model detects illustrate instability periods.}
\label{fig_transform}
\end{figure}

\paragraph{Probabilistic scoring of the incoming data} As described in the previous section, we use a VAR(1) model with 4 variables, scaled to 0.5-second granularity, with training data $\tau$ of 10 minutes for modeling the PMU data-stream. The VAR(1) model is retrained in real-time every 0.5 second and used to predict the next data point in the time series. Once the next data-point is received, a vector-valued error/residual $r_t$ is computed as 
\begin{equation}
 r_t = AX_{t-1} - X_t,
\label{eq_residual}
\end{equation}
where $AX_{t-1}$ is the prediction based on $X_{t-1}$ and $X_{t}$ is the observed measurement. The residuals $r_t$ can then be scored according to the Mahalanobis distance \cite{mah-dist} using noise covariance matrix $\Sigma$ which is learned during the VAR(1) training: 
\begin{equation}
 %\begin{align}
 %\begin{split}
  D_M(\vec{r_t}) = \sqrt{\vec{r_t}^T \ \Sigma^{-1} \ \vec{r_t}}
  \label{eq_dist}
 %\end{split}
 %\end{align}
\end{equation}
The distance of the residuals from the model's noise distribution are unitless and hence transferable across PMU streams and different operating conditions. This metric (given here in the zero-mean case due to de-trending) generalizes the distance in terms of standard deviations for one-dimensional case to the case of multivariate Gaussian distributions.

The analysis of the multivariate distribution can also provide a distance metric related to an individual variable $y_{kt}$ (e.g. voltage alone), but \emph{informed} by the state of other variables. Indeed, a conditional distribution of $y_{kt}$ given the other variables is a Gaussian distribution itself:
% Rather than analyzing the residual distance for the overall vector, we can also measure residuals pertaining to individual variables $y_{kt}$ (such as voltage) in the VAR(1) model. To do so, we determine the conditional distribution of $y_{kt}$ given the other three variables. \textcolor{red}{check this formula:}As given in Equation~\ref{eq_conditionalDist}, the conditional probability is Gaussian with a known covariance matrix, which enables us to compute the Mahalanobis distance.
% \textcolor{blue}{\bf I don't understand the notation in the equation below (commented it out). Why not just put the expression used in the code? I've suggested another expression instead.}
% \begin{equation} 
%  y_{kt} \in y_t | \{y_t - y_{kt}\}) \sim N(\Bar{\mu}, \bar{\Sigma}) 
%  \label{eq_conditionalDist}
% \end{equation}
\begin{equation} 
 r^{k}_{t} \sim N(\Bar{\mu}_{k}, \bar{\sigma}_{k}),
 \label{eq_conditionalDist}
\end{equation}
with mean $\Bar{\mu}_{k} = \Sigma_{k, k^c} \Sigma^{-1}_{k^c, k^c} r^{k^c}_{t}$ and standard deviation $\bar{\sigma}_{k} = \Sigma_{k, k} - \Sigma_{k, k^c} \Sigma^{-1}_{k^c, k^c} \Sigma_{k^c, k}$ that are computed for each variable $k$ and the state of the remaining variables $k^c$.

The incoming data points are provided with the score $S$ equal to the Mahalanobis distance $D_M$ in the case of full multivariate distribution, or to the distance to the mean $\Bar{\mu}_{k}$ measured in standard deviations $\bar{\sigma}_{k}$ for an individual variable $k$. Examples of distances of the individual residuals using the VAR(1) model on the available PMU data is shown in Figure~\ref{fig_transform}. Notice that most of the points lie within the $S=5$ ``$\sigma$'' score, which means that they are considered normal for the current effective model. On the other hand, the larger the value of $S$, the more likely it is to be a severe event. To identify anomalies, we compare the relative score $S$ against a user-specified threshold $T$. When a score larger than the threshold $T$ is found, the VAR(1) model retraining is stopped, as we know that the data set contains anomalous points, and does not describe normal behavior of the system. Our learning algorithm then scores the next $q$ streaming measurement samples by their corresponding residues based on a $q$-step forecast of the VAR(1) process given by $y_{q+t} = A(A(A \dots A(y_{t-1})))$. 
% We assume that the events and anomalies that the detection algorithm is targeting are shorter than $q$ steps ahead. %%%A: I think this sentence is unnecessary.
In our detection approach, we set $q=10$ to detect events of short duration or larger events that progress in this period of time.
%\textcolor{red}{after $q$, we stop.. maybe makes sense to say that we assume that $q$ is sufficiently large to capture the event and that we do not expect two events to occur within $q$ secs}

\begin{figure}%[b] % figure margins 6x8 in
\centering
\figuretitle{Anomaly in raw and in transformed}
\includegraphics[scale=0.65]{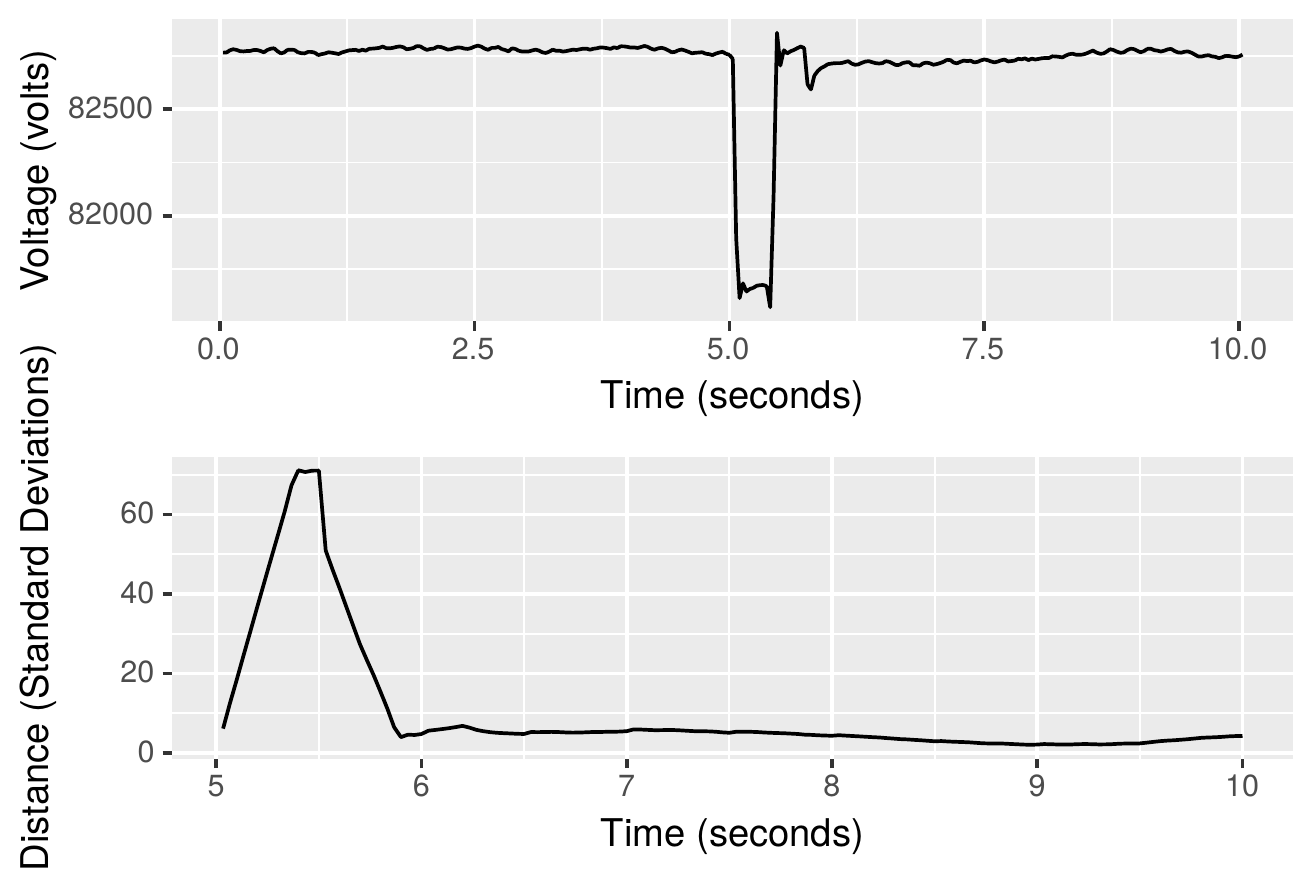}
\caption{A voltage spike of over 1000 volts is detected as an anomaly with a score over 70 standard deviations from the models expected value. 
The quick recovery is seen in the score as well as the voltage data.
}
\label{fig_compare_anomalyV}
\end{figure}

An example of this procedure is given in Figure~\ref{fig_compare_anomalyV} for a voltage anomaly over 10 seconds. The conditional probability is used for scoring the residuals for the 5 seconds following the anomaly. In this anomaly with $S=70$, the deviation of the voltage is over 70 standard deviations away from the expected distribution. 

% The overall sequence of training/re-training and anomaly detection/scoring steps are listed in Algorithm $1$. In this work, we select a $10$ step-ahead ($q = 10$) forecast after threshold crossing to score the anomaly.

% \textcolor{blue}{\bf A: Deep, does this pseudo-code makes things clear for you on what is going on? Certainly not for me. I know you guys added it recently, but I don't see what it adds for a general reader and I'd suggest to remove it altogether. The description in the text seems rather complete and alright to me. An interesting reading could look at the code. Upd: OK, I've commented it out.}
 
% \begin{algorithm}[h]
% \DontPrintSemicolon
% \SetAlgoLined
% \KwData{$Y\gets [y_1 ... y_\tau$] (Data Stream), 
% $T \gets 12$ (Threshold), 
% $p\gets1$ (Lag Terms), 
% %$q \gets 10$ (Classification Window),
% $r \gets 0.5$ (Model Resolution (Sec)), 
% $F \gets 30$ (Collection Frequency (Hz)),
% $\tau \gets 10 * 60 * F$ (Training period (10 Min)), 
% $k \gets \tau$ (Current Data Point) \\
% }
% %\SetKwFunction{SDname}{ProcessPMUStream}
% %\Fn{\SDname{}} { \\
%  $newData \gets$ get\_NewData() \;
%  \While{newData}{
%   $Y_c \gets get\_Coarse(Y[k-\tau:k])$ \;
%   $V \gets get\_VAR(Y_c, lags \gets p)$\;
%   $res \gets V.A \times Y_c[k] - newData$ \;
%   $S \gets get\_Mahalanobis(res, V.\Sigma)$\;
%   \eIf{$S > T$}{
%    $process\_Anomaly(S,V)$\;
%   }{
%   $Y \gets Y.$append$(newData)$\;
%   $newData \gets get\_NewData()$\;
%  }
%  $k++$\;
% }
% %}{\KwRet}
% \caption{Anomaly Detection}% \textcolor{red}{Change Algo Package}}
% \end{algorithm}

\begin{figure}%[b] % figure margins 6x8 in
\centering
\figuretitle{Comparison of Anomaly Detection Mechanisms}
\includegraphics[scale=0.56]{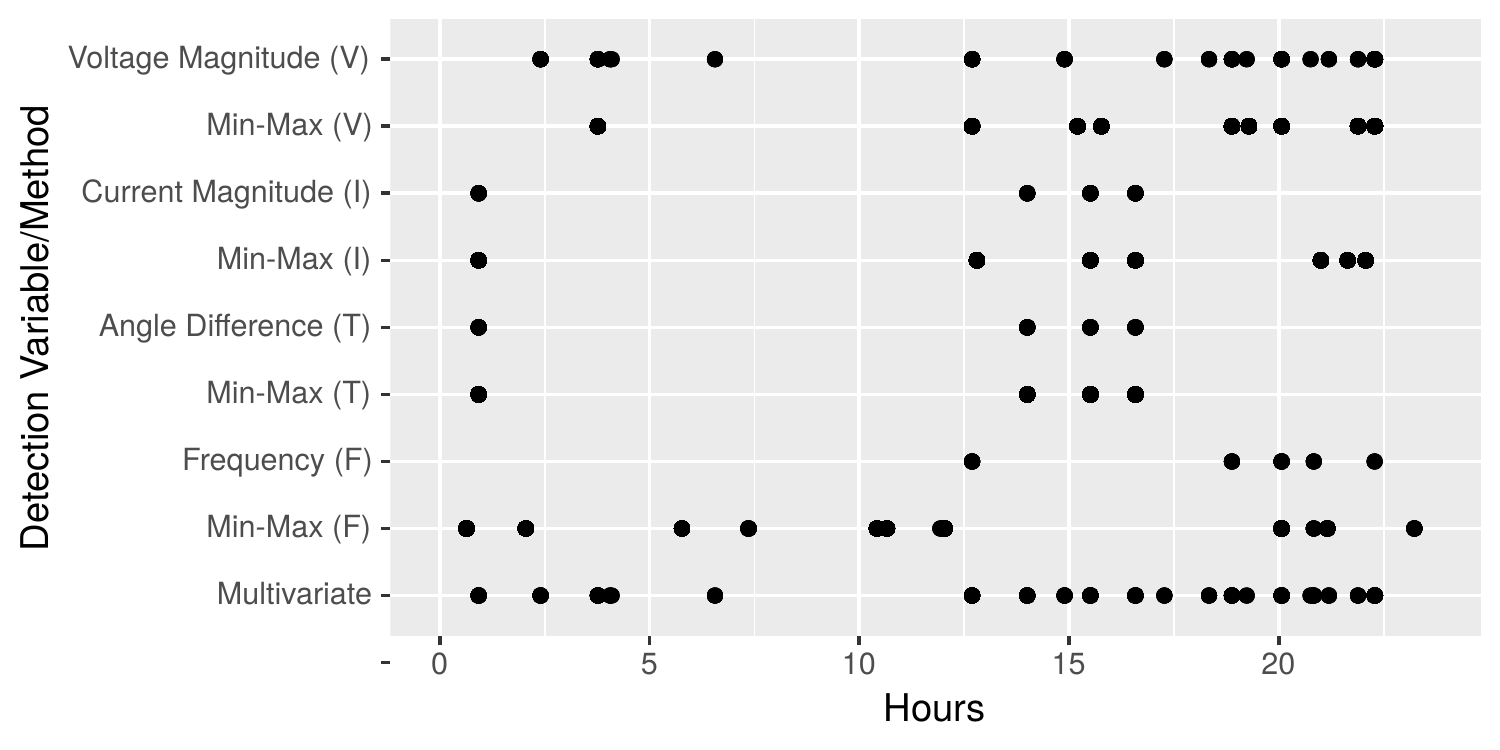}
\caption{This figure illustrates the comparison of the min-max sliding window statistical method applied to each data stream alongside the conditional and Mahalanobis distances of residuals for 1 day. The multivariate detection captures anomalies due to events in individual variables and due to their combined low probability under the learned model.
}
\label{fig_compare_anomaly}
\end{figure}

\paragraph{Multivariate vs. univariate detection} It is worth mentioning that in prior work, anomaly detection is typically based on historical data involving a single variable on a sliding window \cite{pmu_small, pmuOther}. Our approach is more general in utilizing a multivariate statistical linear model.
%We now show that this method is able to perform anomaly detection compared to single variable based tests. 
Figure~\ref{fig_compare_anomaly} compares anomaly detection by our multivariate VAR(1) approach with that based on univariate min-max detection approach on a PMU data stream over 1 day. The min-max algorithm takes the range (maximum-minimum) of a sliding window and determines if it is some standard deviations from the mean of the whole data set. 
Using a 10-second sliding window, there are many overlaps in the conditional probability of VAR residuals as well as the min-max algorithm. For the multivariate approach, we use a threshold of $12$ to identify the start of the anomaly. For the min-max algorithm, the thresholds are set at 3 for voltage magnitude, 4 for current and angle difference, and 6 for frequency. This is in accordance with the National Renewable Energy Laboratory's PMU anomaly detection guidelines \cite{nrel_long} which suggest that the use of multiple techniques is useful to detect changes based on historical PMU data. It is clear from Figure~\ref{fig_compare_anomaly} that our multivariate approach is able to identify almost all anomalies that are detected in different single variable statistical analysis. Crucially, it is able to overcome the fact that certain anomalies detected by a single-variable based test are missed entirely by tests based on some other variable. This combined with the automatic re-training during normal operating regime makes our framework seamless yet accurate.

\begin{figure}%[htp]
\centering
\subfloat[5 Standard Deviations]{\includegraphics[scale=0.52]{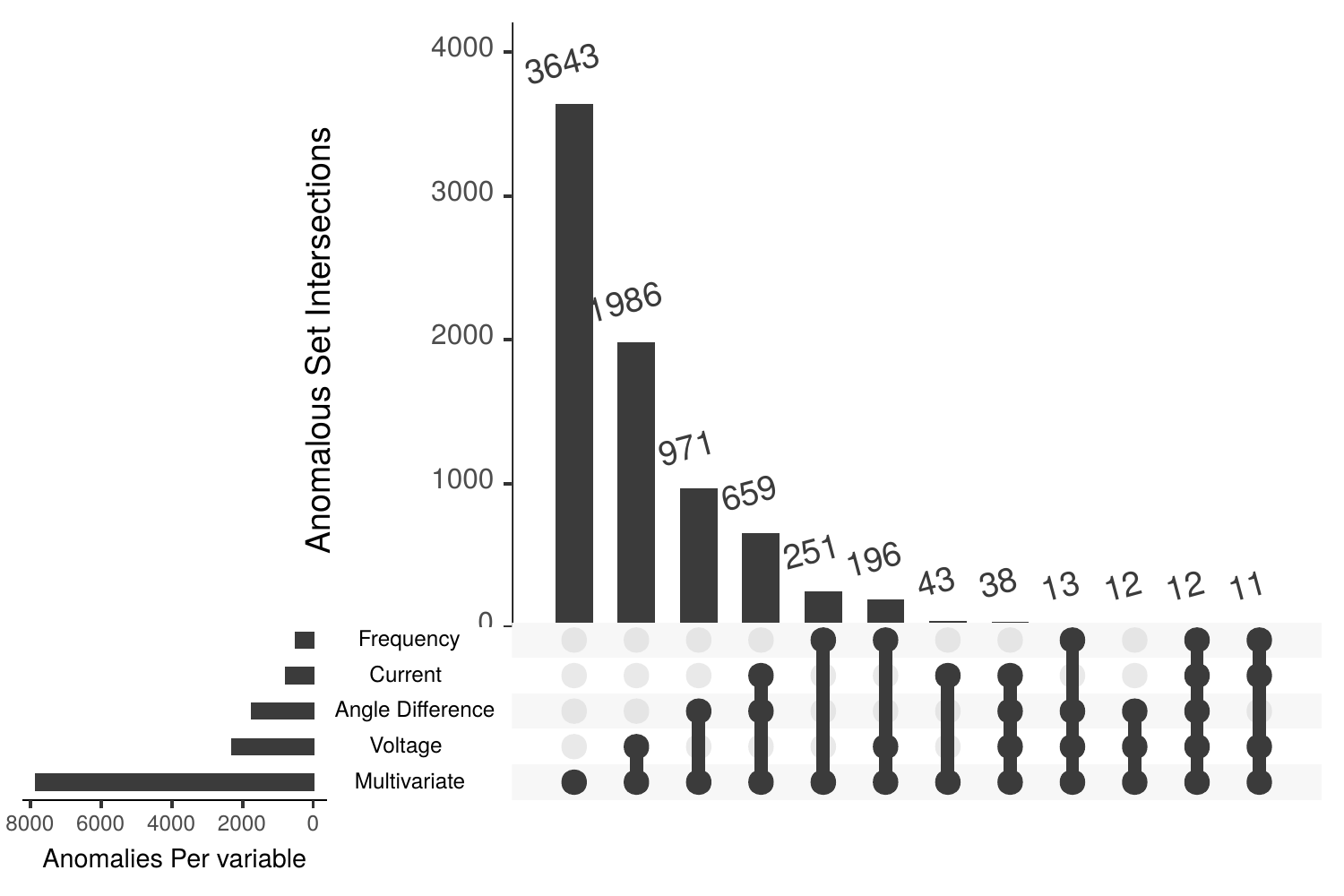}}\\
\subfloat[15 Standard Deviations]{\includegraphics[scale=0.52]{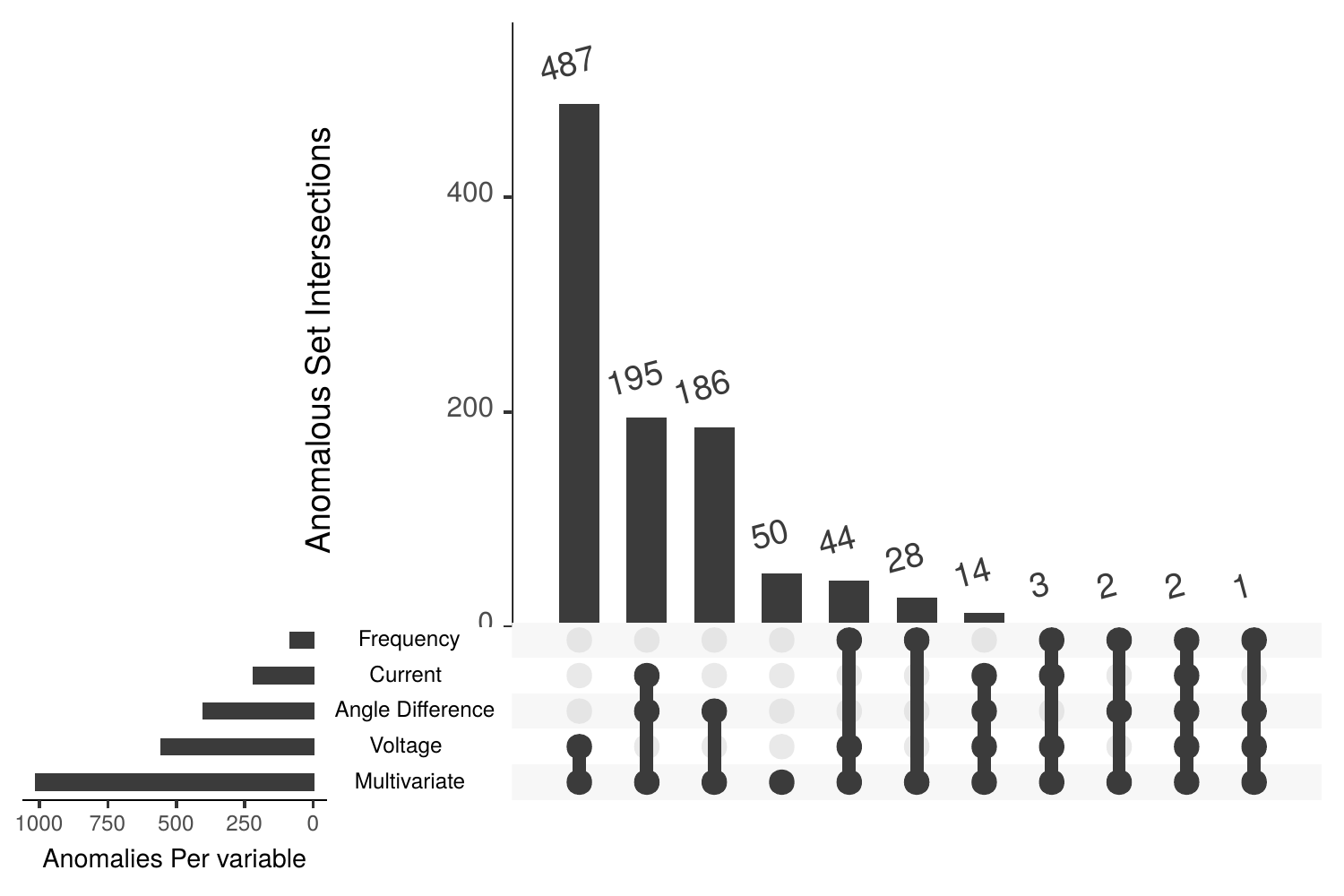}}
\caption{Number and co-occurrence of anomalies in different variables detected on 87 days of historical data scored by Mahalanobis distances on data residuals. $T=5$ and $T=15$ ``standard deviations'' are used as thresholds to detect anomalies. The multivariate Mahalanobis distance shows an anomaly when the conditional probability distance for each variable does. Plots are created using UpSetR package \cite{upset}.}
\label{fig_upset5}
\end{figure}

\paragraph{Selection of the detection threshold} To demonstrate the dependence on the threshold $T$, we consider the entire data-stream of $87$ days and report results on anomaly detection by multivariate as well as individual variables in Figure~\ref{fig_upset5} for two thresholds (a) $T= 5$, and (b) $T= 15$. In either case, each detected anomaly labeled based on the set of individual variables scores and/or multivariate score that led to its discovery. Setting a detection threshold of $T$ = 5 standard deviations results in a total of just under 8000 anomalies. Observe that the most common detection of anomalies corresponds to detection through the multivariate case. Voltage anomalies are the second most occurring anomaly type, but they are also detected by the multivariate test. A threshold corresponding to five standard deviations may be too low as in the large amount of data it would also capture the low-probability deviations that correspond to normal events under the learned model. Setting a detection threshold of $T = 15$ in Figure~\ref{fig_upset5} (b), reduces the number of false positives but may potentially miss some low impact events. The trend remains similar in both cases: the multivariate score is enough to detect the score in all other variables. In practice, a threshold selection should be done by the operator using randomized tests with labeled data of events of importance to the system. In this threshold setting scheme, one could imagine an-operator assisted procedure where a human may introduce some feedback to fine-tune the threshold since the events are detected in real-time. In the next step, we move to the crucial step of anomaly classification of detected events.

\section{Classification of Anomalies}
\label{sec-classification}
It is desirable to provide information useful for operators once an event is detected. Specifically, a tool that is able provide an automatic classification of the anomaly into predefined classes can provide quick insight into the operation of the system. This would enable the operator (or automated system) to take fast and informative control action. 
Another requirement for the classifier is to be as interpretable as possible: the operator should be able to look up why a certain event has been classified into a given class.  

From the detection results, it is clear that among single variable based detection tests, voltage anomalies are the most frequent anomaly type. To describe and demonstrate our classification task, we focus on assigning the voltage anomalies into a set of common occurrence classes. The PMU data available to us is unlabeled, and hence for the sake of illustration of our methods, we resort to shape-based but intuitive classes. We have manually labeled 750 voltage events into 4 categories using raw data of the voltage signal: voltage spikes, prolonged voltage drops, step up and step down events typically caused by load tap changes and miscellaneous oscillatory events. In real-world application, these classes should be defined based on the domain expertise of the operators. 

\begin{figure}%[ht]%[b] % figure margins 6x8 in
\centering
\figuretitle{Decision Tree Voltage Classification}
\includegraphics[scale=0.4367]{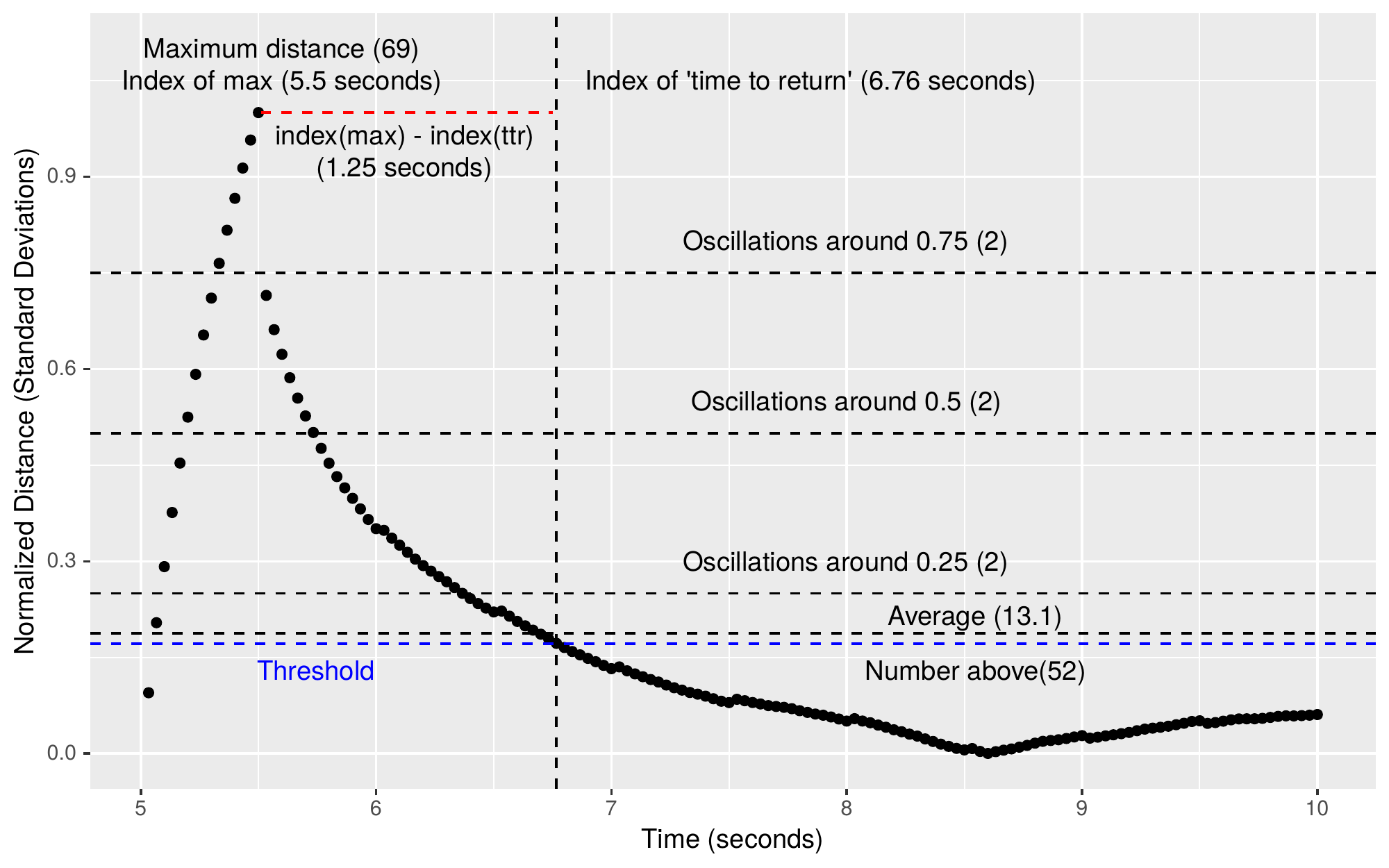}
\caption{A representative anomaly of length $5$ seconds with annotated features is shown. The detection threshold is shown in blue. The Deciles are not shown due to space constraints.
}
\label{fig_features}
\end{figure}

To enable automatic classification, data features are generated from the data residuals, using 5 seconds of data following anomaly detection. The extracted features include:
\begin{itemize}
 \item Maximum distance
 \item Average distance
 \item Number of points above the threshold
 \item Decile Ranks 1-9
 \item Index of maximum distance
 \item \# of oscillations around the 0.25, 0.5, and 0.75 levels
 \item Return from anomaly index. (Index of value such that future residuals are all below the anomaly threshold)
 \item The difference of indices of maximum distance and return from anomaly
\end{itemize}
For elucidation, a representative anomaly of length $5$ seconds with annotated features is shown in Figure~\ref{fig_features}. In practice, additional features can be generated using automatic feature extraction schemes or using domain knowledge .

% \textcolor{blue}{\bf A: the precise meaning of each and every feature is not very clear. Ideally, one would like to take a picture like Fig. 8 and schematically illustrate all of the features as a picture on top of the signal. Would that be possible?}

\begin{figure}%[ht]%[b] % figure margins 6x8 in
\centering
\figuretitle{Decision Tree Voltage Classification}
\includegraphics[scale=0.575]{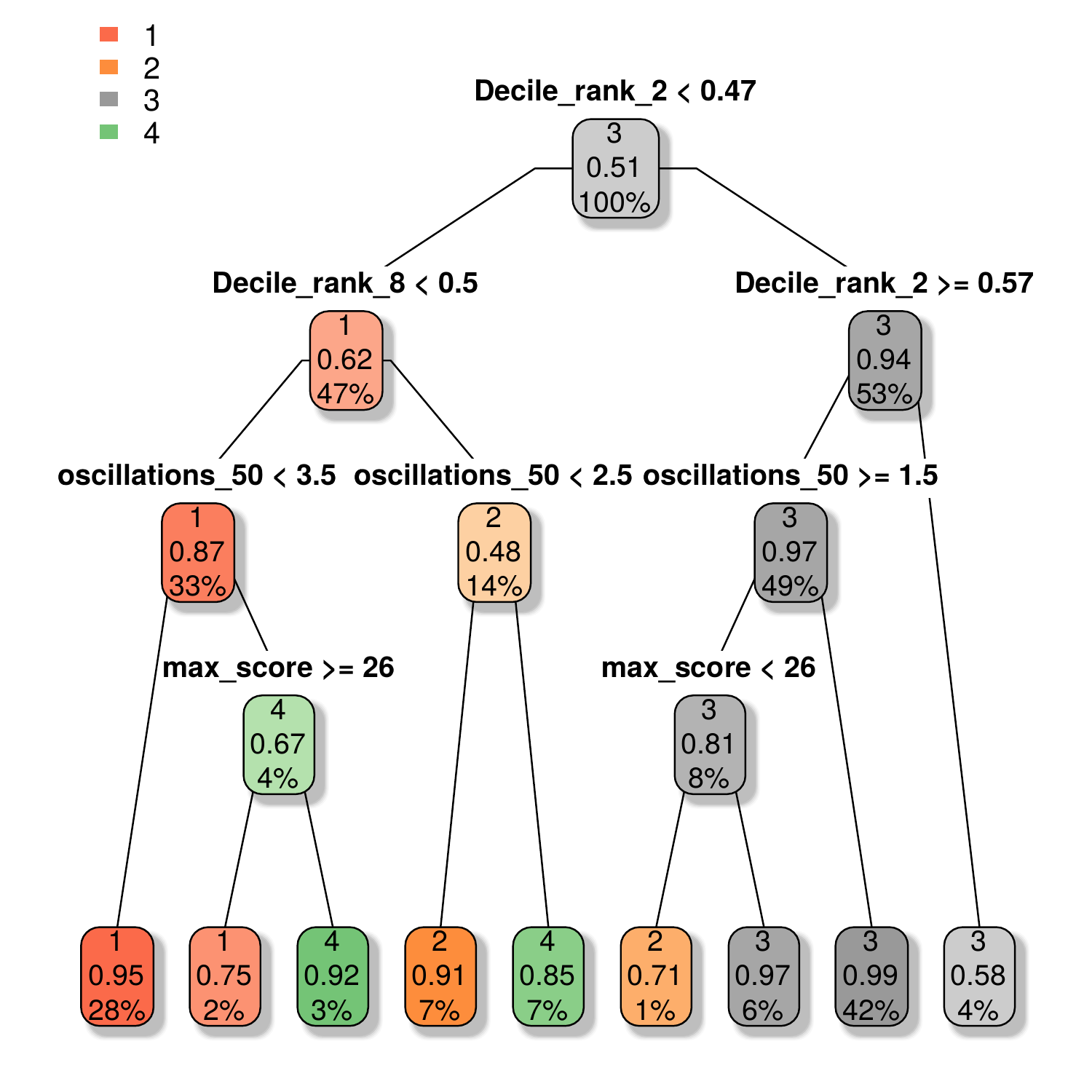}
\caption{Decision tree using automatically selected features: decile ranks 2 and 8, the maximum distance and the oscillations around the mean distance. In each node, the top number corresponds with color to the prediction at that node. The middle number represents the accuracy of that prediction at that node and the bottom number represents the percent of training samples that fall into that node. 
}
\label{fig_decisionTree}
\end{figure}

Our classification scheme is based on a supervised decision tree \cite{cart84} %\textcolor{blue}{\bf insert citation} 
based on the above features. The decision tree partitions regressively on features using the algorithm \cite{rpart} and the R library \textit{rpart} \cite{rpart_r}. Among others, decision trees have good interpretability as they offer operators the exact sequence of feature values that led to an anomaly being classified into a particular class. In practice, this can help in the selection or removal of specific features based on the accuracy over labeled training data. The regressive partitioning algorithm in our decision tree selects a subset of features that give the most accurate classification: decile ranks 2 and 8, the maximum distance and the oscillations around the mean distance.
Figure~\ref{fig_decisionTree} shows the view of the decision tree used for our data. Using 10-fold cross-validation, we present the accuracy of classification by our decision tree. To benchmark its performance, we compare it with black-box neural network and k-means clustering-based classification using both features extracted from the data and raw data in Table~\ref{tab:comp}. Despite its simplicity and interpretability, the decision tree approach has an accuracy of 93.3\%, which is on par and even higher than that of other learning schemes.

\begin{table}[ht]
\centering
 \caption{Performance comparison of different classification methods}
 \begin{tabular}{| l | c | }
 \hline
 {\bf Classification Approach} & {\bf Accuracy} \\ \hline
 Decision Tree (features) & 93.3\% \\ \hline
 Neural Network 0 hidden layers (features) & 91.1\% \\ \hline
 Neural Network 10-100-10 hidden layers (features) & 91.7\% \\ \hline
 Neural Network 10 hidden layers (raw data) & 86.0\% \\ \hline
 K-means Clustering (features) & 90.9\% \\
 \hline
 \end{tabular}
 \label{tab:comp}
\end{table}

\section{Discussion and conclusions}
\label{sec-discuss}

Large quantities of synchronized data from phasor measurement units have created a need for a statistical learning based framework to process, model and evaluate the data to gain insight and understanding of the electric power system. The requirements for such a framework include: (a) \emph{generality}: the scheme should not be system-specific and should be transferrable and deployed on a large number of infrastructure systems; (b) \emph{interpretability}: each step, be it model learned, anomaly detected, or classified, should be easily interpretable by an operator to better inform the response decisions; (c) \emph{modularity}: each of the components could potentially be replaced by other schemes if desired; (d) \emph{guarantees}: well-defined theoretical and empirical error measures provided with each step of the framework; and (e) \emph{computational efficiency}: the methods should operate on streaming data, perform real-time computations, and provide early detection, classification, and explanation of the events.

This paper shows a general real-time approach that satisfies all the criteria listed above. It utilizes an effective physics-motivated linear vector autoregressive model learned from a PMU data stream. We presented the general data-driven procedure that can be used to select hyper-parameters of the learning scheme in a principled way. The value of the effective model learned is shown through two important applications: categorization of anomalies detected using well-defined probabilistic scores according to the learned model; and classification of anomalies using interpretable decision trees that also show a very good level of accuracy. 

One of the advantages of the designed anomaly detection scheme consists in the choice of the invariant representation of the signals in terms of probabilistic scoring according to the effective model learned on the fly. This ensures the transferability of methods to very general settings. As far as  classification is concerned, given that most of the data is unlabeled, in future work it is important to move to unsupervised approaches such as clustering to be able to classify unlabeled data corresponding to different variables automatically.

One important future direction is to use the full power of the learned effective model to see the signs of endogenous events or cyber attacks \cite{lokhov2016detection} before they occur. Inherently, it is impossible to predict external perturbations, but it is possible to predict the response of the system to exogenous perturbations and to take corrective actions if the response is dangerous for the stability of the system. We anticipate that for many such scenarios, the information should be contained in the dynamic state matrix of the effective model learned prior to the event. %This path will be explored in more details in future work.

This work focused on establishing data processing, anomaly detection and classification at the level of a single PMU with multi-variable stream. Moving forward, it is important to use the inferred information for network-wide tasks, such as localization of anomalies \cite{8718345}. With the overwhelming quantity of data transmitted in modern WAMS, and limited comunication network infrastrucure, it is desirable for data to be processed locally on each PMU in the way described in this study; and that only labeled information on the detected events with time stamps could be sent to the central server for performing network-wide analysis from a partial PMU coverage \cite{lokhov2018cdc}.%, deka2017state}. %The development of such a protocol will avoid overloading the communication system, and is left for future work.

Last, our short term goal will be to release code associated with the methods developed in this study with a visualization component, which would enhance the interpretability of the results for grid operations.

\vspace{-0.15cm}
\section{Acknowledgments}
We are grateful to Scott Backhaus, Trevor Crawford and Alaa Moussawi for fruitful discussions and for their insights in formulating various elements of the developed framework.
           %
%%%%%%%%%%%%%%%%%%%%%%%%%%%%%%%%%%%%%%%%%%
\vspace{-0.15cm}
\bibliographystyle{IEEEtran}
\bibliography{references.bib}{}


\end{document}